\documentclass[conference,a4paper]{IEEEtran}
\usepackage[utf8]{inputenc}
\usepackage[english]{babel}

\usepackage{cite}
\usepackage{amsmath,amssymb,amsfonts}
\usepackage{algorithmic}
\usepackage{graphicx}
\usepackage{textcomp}
\usepackage{xcolor}
\usepackage{soul}
\usepackage{enumerate}
\usepackage{comment}

\usepackage{subcaption}
\usepackage{multirow}
\usepackage{tabu}
\usepackage{float}
\usepackage[inline]{enumitem}

\usepackage[autostyle]{csquotes}
\usepackage[utf8]{inputenc}

\definecolor{boristext}{rgb}{0.3, 0.36, 0.88}
\definecolor{boriscomments}{rgb}{0.83, 0.0, 0.0}
\definecolor{davidcomments}{rgb}{0.0, 0.0, 0.83}

\IEEEoverridecommandlockouts

\begin{document}

%\title{Group Creation and Traffic Scheduling for Coordinated Spatial Reuse in Multi-AP WLANs}
%\title{Group Creation and Scheduling for Coordinated Spatial Reuse in Multi-AP WLANs}
%\title{Coordinated Spatial Reuse for Wi-Fi 8: Finding Compatible APs and Group Scheduling}
\title{Multi-AP Coordinated Spatial Reuse for Wi-Fi 8: Group Creation and Scheduling}
%\author{David Nunez, Malcolm Smith, Boris Bellalta}
%\date{04/15/2022}

\author{
\IEEEauthorblockN{David Nunez$^{\star}$, Malcom Smith$^{\flat}$, and Boris Bellalta$^{\star}$\vspace{0.05cm}
}
\IEEEauthorblockA{$^{\star}$\emph{Wireless Networking @ Universitat Pompeu Fabra, Barcelona; $^{\flat}$\emph{Cisco Inc., USA}}}
%\IEEEauthorblockA{$^{\flat}$\emph{Cisco Inc., USA}}
\thanks{D. Nunez and B. Bellalta were supported in part by grants Wi-XR PID2021-123995NB-I00 (MCIU/AEI/FEDER,UE), and Cisco. The authors thank the comments and inputs received while preparing this work by Prof. S. Avallone and Dr. P. Imputato.}
}

\maketitle

% ---------------------------
% ---------------------------
% ---------------------------
% ---------------------------

\begin{abstract}

Multi-Access Point Coordination (MAPC) will be a key feature in next generation Wi-Fi 8 networks. MAPC aims to improve the overall network performance by allowing Access Points (APs) to share time, frequency and/or spatial resources in a coordinated way, thus alleviating inter-AP contention and enabling new multi-AP channel access strategies. 
This paper introduces a framework to support periodic MAPC transmissions on top of current Wi-Fi operation. We first focus on the problem of creating multi-AP groups that can transmit simultaneously to leverage Spatial Reuse opportunities. Then, once these groups are created, we study different scheduling algorithms to determine which groups will transmit at every MAPC transmission. Two different types of algorithms are tested: per-AP, and per-Group. While per-AP algorithms base their scheduling decision on the buffer state of individual APs, per-Group algorithms do that taking into account the aggregate buffer state of all APs in a group.
%
%to schedule transmissions for real-time traffic, through the use of cooperative multi-AP networks with coordinated time division multiple access (c-TDMA) and coordinated-TDMA with spatial reuse (c-TDMA/SR) schemes. \bcom{I would not mention the name of the algorithms here.} We propose a method to assess the spatial reuse compatibility between Access Points, and an algorithm to create groups of SR-compatible APs.
%
Obtained results---targetting worst-case delay---show that per-AP based algorithms outperform per-Group ones due to their ability to guarantee that the AP with a) more packets, or b) with the oldest waiting packet in the buffer is selected.%, which at the end, it is the key factor to minimize the worst-case delay. 
%Results show \st{that the algorithms proposed for c-TDMA/SR perform better in terms of throughput, delay, and slot occupancy than the ones proposed for c-TDMA, in most cases}. \bcom{Well, this is obvious. What it is important is what we 'learn' about the group creation, and how group vs individual algorithms perform, as well as between throughput and delay ones}.  

\end{abstract}

\begin{IEEEkeywords}
Multi-AP Coordination, Coordinated Spatial Reuse, Wi-Fi 7, Wi-Fi 8, IEEE 802.11be, IEEE 802.11bn, WLANs. 
\end{IEEEkeywords}

% \tableofcontents

\section{Introduction}

%\textcolor{red}{Font!!!!!}

Nowadays, the consumption of high-bandwidth and real-time applications constitutes a massive challenge for operators and network companies to deliver these contents to end users. Video streaming is becoming more popular than ever. Only in the first half of 2021 it represented the  53.72\% of all Internet downlink traffic \cite{Internet_traffic_Sandvine}. Beyond traditional video streaming consumption, cloud gaming \cite{carrascosa2022cloud}, virtual and augmented reality (VR/AR) \cite{zhao2021virtual} are rapidly becoming more and more popular, hence further contributing to increasing the demand of interactive and delay-sensitive contents. %As an example of such raising interest, the worldwide market for VR/AR headsets grew 92.1\% in 2021 with shipments reaching 11.2 million units \cite{AR-VR_IDC}. %, with its demand  increasing over time. %, allowing many users to play games remotely. % It is expected that the number of users worldwide will reach 349.4 million by 2025, compared to 60.6 million users in 2021 \cite{Cloud_Gaming_Research}. Besides, virtual and augmented reality (VR/AR) industry face significant changes every year. The worldwide market for headsets of these technologies grew 92.1\% in 2021 with shipments reaching 11.2 million units \cite{AR-VR_IDC}. 

% Several physical layer enhancements were also proposed for the 11be amendment and other technologies like multi-link operation (MLO) \cite{naribole2020simultaneous,naik2021can,lopez2021ieee,park2021latency,carrascosa2021experimental} are being developed to improve also the link layer.

%  At this moment, network manufacturers already commercialize Wi-Fi devices based on IEEE~802.11ax amendment (aka Wi-Fi 6) \cite{bellalta2016ieee} and research companies are focused on IEEE~802.11be (11be) amendment~\cite{draft11be}, which represents the foundations of Wi-Fi 7.

% The list of the most popular video apps was headed by Youtube at \#1 with 14.61\%, Netflix at \#2 with 9.39\%, Facebook Video at 4.20\%, and TikTok at 4.00\% of total app traffic. Disney+ is rapidly moving up the ranks at \#15, surpassing Amazon Prime at \#16 \cite{Internet_traffic_Sandvine}.

% Due to a large volume of data is periodically sent from a local device to servers and that most of the processing of the content occurs at remote ends (proprietary servers are often far from users), cloud gaming constitutes a big picture when it needs to illustrate a high-throughput and low-latency service. 

To get over that, Wi-Fi networks are constantly evolving to address not only the high requirements of these applications in terms of throughput and/or latency, but also the increasing number of users and the traffic volume on Internet. Access Point (AP) densification (i.e., covering the same area with a high number of APs) has been the natural response to cope with such a situation. This approach allows stations to benefit from high Signal to Noise (SNR) levels, as they are close to their serving APs, resulting in the use of high transmission rates. However, when the number of co-located APs is high, the limited number of frequency channels may result in detrimental high contention and interference levels, as well as affecting the ability of Wi-Fi networks to provide a reliable service. Improving this situation is set as a requirement for future Wi-Fi 8 by the 802.11 Ultra-High Reliable (UHR) Study Group   \cite{UHRobjectives}.

A solution to mitigate the high contention levels in dense Wi-Fi deployments is to coordinate transmissions from the set of overlapping APs. To support such an objective, the Multi Access Point Coordination (MAPC) framework \cite{AdrianGarciaSurvey,EvgenySurvey,TXOPsharingPaper} was initially included as part of the IEEE~802.11be (11be) amendment~\cite{draft11be} candidate features, although its development has been postponed for the future Wi-Fi 8 amendment ---likely to be named as IEEE 802.11bn--- leaving Multi-link Operation~\cite{lopez2022multi,carrascosa2022experimental} as the key and most disruptive feature of IEEE 802.11be.  
%Several schemes have been proposed to be used in along with MAPC: coordinated Time Division Multiple Access (c-TDMA), coordinated Spatial Reuse (c-SR) and coordinated Orthogonal Frequency Division Multiple Access (c-OFDMA) are the most suitable candidates to be included in 802.11be, or future amendments, due to their simplicity. In this regard, 
%, a must is to develop efficient MAPC schedulers to deal with different traffic categories and the need of coexistence with legacy networks. 
%A solution to mitigate this issue could be based on scheduling coordinated transmissions. Exchanging information periodically between coordinated devices allows more accurate scheduling decisions. To perform these tasks, multi access point coordination (MAPC) \cite{TXOPsharingPaper,AdrianGarciaSurvey,EvgenySurvey} is being developed as part of the group of candidate features for IEEE~802.11be (11be) amendment~\cite{draft11be} and several schemes have been proposed to be used in it. Among them coordinated Time Division Multiple Access (c-TDMA), coordinated Spatial Reuse (c-SR) and coordinated Orthogonal Frequency Division Multiple Access (c-OFDMA) are the most suitable candidates to be included in 11be due to their simplicity. In this regard, the research community is mainly focused on developing efficient schedulers to deal with the problem of scheduling different traffic categories and the need of MAPC coexistence with legacy networks. 
MAPC allows APs to share time, frequency and/or spatial resources in a controlled manner, alleviating Overlapping Basic Service Set (OBSS) contention, and enabling the implementation of WLAN-level scheduling mechanisms. However, there are still several challenges to solve. Among them, we need (i) to design a protocol for coordinating the transmissions from the set of overlapping APs; (ii) a mechanism to decide which of the overlapping APs are compatible to transmit at the same time by, for instance, leveraging Spatial Reuse (SR) opportunities; and (iii) a mechanism to decide which of those compatible APs are allocated to each coordinated transmission. %In this paper we cover these three challenges for the case in which Spatial Reuse is considered by the MAPC framework to support multiple simultaneous transmissions. The presented MAPC framework

In this paper, to deal with these aforementioned challenges, we present a framework to support MAPC on Wi-Fi networks. It combines $i)$ periodic dowlink `MAPC transmissions' that are able to leverage Spatial Reuse opportunities when possible, with $ii)$ un-coordinated CSMA/CA `breathing' periods in between to account for other downlink and uplink traffic. %\bcom{Too many details in this paragraph} We assume the existence of a controller-based solution to support c-TDMA/SR transmissions in the periodic slots, splitting the process in two stages: $i)$ creation of compatible groups of APs (i.e., APs that can perform simultaneous transmissions by leveraging SR), and $ii)$ group scheduling at every slot using the AP's buffer state as metric. On the other hand, un-coordinated transmissions occur in between coordinated slots to allow either AP and stations to deliver their best-effort traffic, as well as any management and control traffic exchange between stations and APs. We also detail an over-the-air protocol to coordinate channel access between the overlapping APs. Two types of traffic scheduling algorithms are also presented: $i)$ based on individual AP requirements, and $ii)$ based on the requirements of a group of compatible APs. Their performance assessment is one of the key contributions of this paper.
In detail, the main contributions of this paper are:
\begin{enumerate}
    \item A MAPC framework, to leverage both coordinated and uncoordinated transmissions on top of a multi-AP WLAN.
    \item A low-complexity algorithm to build groups of compatible APs able to transmit simultaneously by leveraging Spatial Reuse opportunities.
    \item Different scheduling algorithms to select the groups of APs that will be scheduled in each MAPC transmission. Two types of algorithms are considered: per-AP and per-Group algorithms.
    \item Insights on how to configure several key parameters used by the group creation algorithm, and on the performance of the per-AP vs per-Group scheduling algorithms.
\end{enumerate}

This paper extends the work done in \cite{TXOPsharingPaper}. Novel aspects in this paper include the definition of practical algorithms for the creation of compatible groups and group scheduling under finite load conditions. 

%This paper \st{shows a model} presents a framework that combines both carrier sense multiple access with collision avoidance (CSMA/CA) and MAPC transmissions, and is focused on a framework that periodically schedules slots for real-time traffic using c-TDMA and when possible, applies c-SR to boost these coordinated transmissions. We propose a controller-based solution to support these c-SR transmissions in those slots, splitting the process in two stages: group creation, and scheduling taking into account buffer information. On the other hand, CSMA/CA transmissions occurs in between as part of the coexistence with legacy devices. We also present a control-plane mechanism to exchange information between APs through a wired backhaul, as well as a wireless framework to reserve the channel and share other parameters between APs. Several algorithms are presented, aiming to compare if individual or group based scheduling are the best way to attain a certain goal. The assessment of them in different load conditions is one of the key contributions of this paper. 

\section{Related Work}

Currently, there are only a few works that delve into MAPC to leverage Spatial Reuse opportunities, and most of the information available are still directly coming from TGbe documents. In~\cite{MentorConsiderations0590r5}, the process to transmit in coordinated Spatial Reuse (c-SR) mode is split into three phases and the authors investigate several operation issues, such as information exchange about transmission power levels (one-way or bidirectional), path loss, and block acknowledgment. In addition, authors  in~\cite{MentorSR1534r1,MentorSR0107r1,MentorSR0576r1} provide simulation results showing the potential performance gains of c-SR. For example, the work in~\cite{MentorSR0107r1} showcases some of the benefits of using c-SR compared with the default Enhanced Distributed Channel Access (EDCA) mechanism. Besides, in~\cite{MentorSR1534r1}, the gains of c-SR are two times higher than the legacy systems. Similarly, the authors in \cite{MentorSR0576r1} make a comparison between c-SR and coordinated Orthogonal Frequency Division Multiple Access (c-OFDMA), showing that the throughput for the former exceeds (twice in some cases) the latter. The work in~\cite{WoojinCoOFDMAframework} exhibits a novel transmission scheme for 11be networks, utilizing the concept of multi-AP c-OFDMA. The author shows that the c-OFDMA scheme effectively allows APs to increase the number of transmission opportunities, achieving a higher throughput than DL OFDMA in IEEE 802.11ax. Finally, the authors in \cite{Yuto_kihira_cSR_Qlearning}, propose a scheme to identify the interference-free APs and a method to reduce the amount of shared information between coordinated devices using Q-learning. 
To the best of our knowledge, none of the works published so far deal with the scheduling process in MAPC networks and also mechanisms to create groups of compatible APs that employ c-SR scheme. Thus, this work explores alternatives to cover these open challenges and also evaluates through simulations the results when each of the proposed algorithms is employed.

% ---------------------------------------------__
% ---------------------------------------------__
% ---------------------------------------------__
% ---------------------------------------------__

\section{Multi-AP Coordination}\label{section:MAPC}

%\bcom{How viable is to replace slots by MAPC transmissions?}

%\bcom{To explain (Boris): We should relax the idea of slots: all is CSMA/CA, but from time to time (periodically, although it could follow a different pattern) the CC decides to schedule a coordinated transmission. Those coordinated transmissions may use favorable EDCA parameters to guarantee almost always win channel contention. In the paper, we assume that they always win the contention anyway.}

\begin{figure*}[t!!!!!]
    \centering
    \includegraphics[scale=0.7]{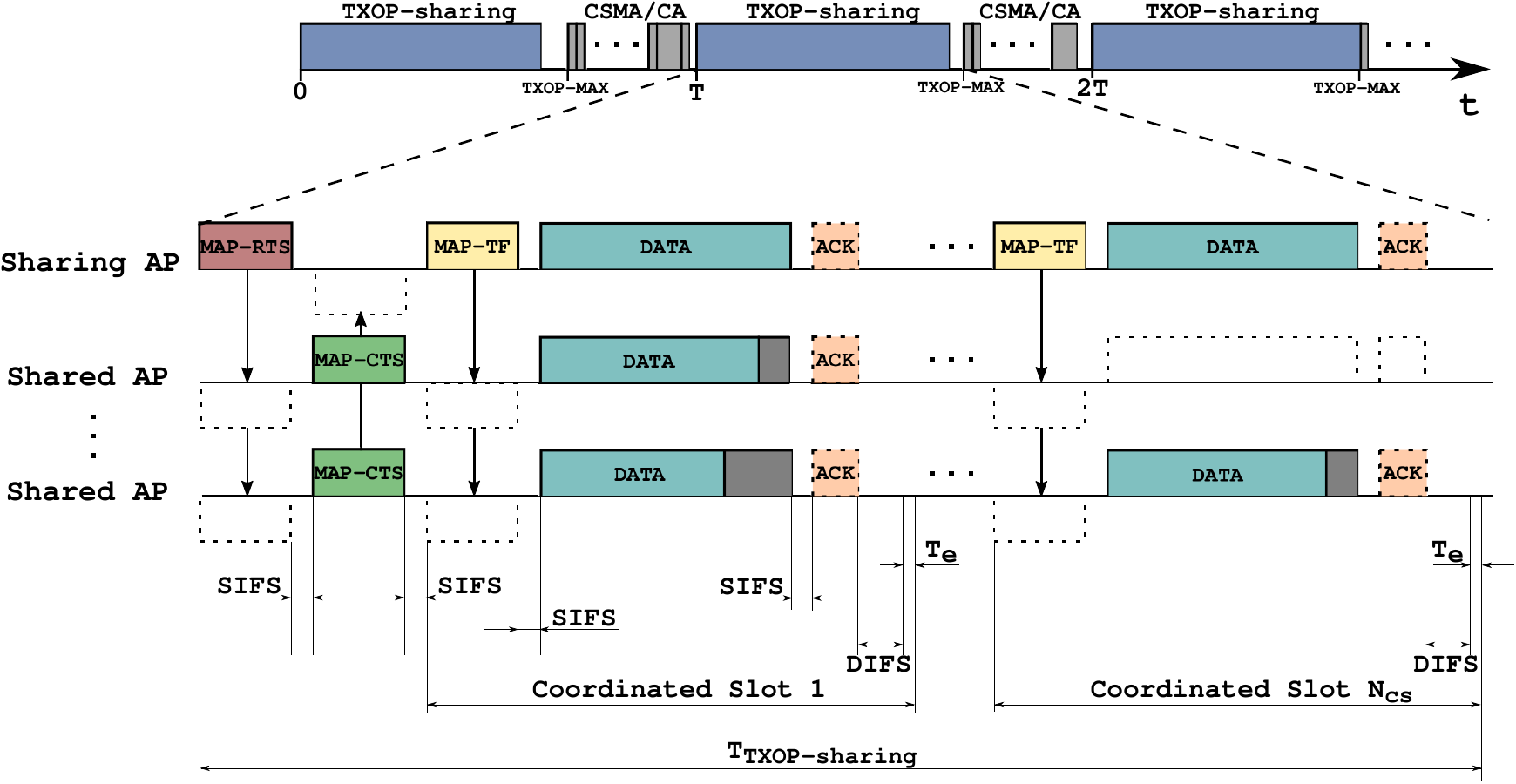}
    \caption{MAPC framework operation. Periodic slots start at every $T$ ms with un-coordinated transmissions in between. The TXOP-sharing period is divided into coordinated slots, allowing both c-TDMA and/or c-TDMA/SR transmissions.}
    \label{Fig:c-TDMA/SR}
\end{figure*}

%\subsection{Description}\label{subsection:Decription}

To support MAPC transmissions, we consider all APs share information with a central controller (CC\footnote{This function could be implemented on a device like an AP, but for a more realistic approach and to avoid complexity on access points, we consider the existence of an external central controller.}). The CC is responsible for computing and periodically informing transmission parameters, as well as the most suitable subset of transmitting APs for every Transmission Opportunity (TXOP). APs and CC exchange most of the information through a low-latency and high bandwidth wired backbone. 

One of the APs acts as the MAPC transmission initiator, while the other APs will simply follow the received indications. We will refer to the AP initiating the shared transmissions as the Sharing AP, and to the rest of APs as the Shared APs. Therefore, the Sharing AP is in charge to initiate the transmission, reserving the channel, and inform which other APs will participate in the TXOP, including the parameters to do so. In this paper we assume all APs are within the coverage area of the Sharing AP.

The operation of the proposed MAPC framework is shown in Fig.~\ref{Fig:c-TDMA/SR}. It allows several APs to share a TXOP opportunity. In the proposed framework, MAPC transmissions are scheduled periodically every $T$ ms with uncoordinated (default CSMA/CA) transmissions in between. A MAPC transmission is initiated when the Sharing AP accesses the channel and sends a MAP request-to-send (MAP-RTS) frame for channel reservation.\footnote{Favourable EDCA parameters can be used to almost always guarantee that MAPC transmissions win access to the channel as earlier as possible.} If no collision occurs,\footnote{Although not extrictly necessary, the use or Restricted Target Wake Time (R-TWT) \cite{draft11be} may provide extra protection to MAPC transmission.} the Shared APs reply at the same time with a MAP clear-to-send (MAP-CTS) frame. The latter is also useful to inform stations or legacy devices about the ongoing MAPC transmission, so it avoids unwanted collisions with hidden devices. At this point, the Sharing AP assumes that all neighboring devices have properly set their network allocation vector (NAV), and so the MAPC transmission will not be disturbed until it ends. Since it is the Sharing AP who shares the TXOP with the other APs, we will use both MAPC transmission and TXOP sharing transmission terms indistinctly. 

Once the Sharing AP accesses the channel, it splits the gained TXOP in one or more temporal slots, to which we refer as coordinated slots. One or more APs can be allocated in each coordinated slot. To do that, the Sharing AP sends a MAP trigger frame (MAP-TF) to allocate the next coordinated slot to one or more APs, including also the settings to use, such as the modulation and coding scheme (MCS) to be used by the Shared APs. 
%the best settings that include \st{total bandwidth} the channel width, MCS, that Shared APs will use in the upcoming transmission. 
When a coordinated slot is allocated to a single AP, the TXOP is shared following a traditional time division multiple access (TDMA) scheme. Otherwise, if several APs are allocated to the same coordinated slot, TDMA is enhanced with Spatial Reuse.

%In order to allocate a coordinated slot to a given group of APs, 
To create the groups of compatible APs that can be allocated to the same coordinated slot, the CC needs to know the received signal strength indicator (RSSI) at the stations from all APs. Due to the symmetry of the channel, we consider that this information is obtained from the opposite path, i.e., by measuring the received power from uplink frames (either data or ACK\footnote{We assume ACKs are correctly received in all cases.} frames). Each AP stores the RSSI of all overheard stations, sharing this information with the CC through the wired backbone.

The uncoordinated CSMA/CA period after each MAPC transmission can be used by any AP and station to transmit either downlink or uplink traffic following default Wi-Fi operation.

\section{c-TDMA/SR scheduling}

%\bcom{To check if it can be moved before the system model. If psosible, that would be the best.}

The MAPC scheduling is split into two main stages: the \textbf{First Stage} includes the creation of the SR-Compatible groups\footnote{Groups of APs that can transmit simultaneously using SR.} based on the signal-to-interference-plus-noise ratio (SINR) between all AP and station pairs, and the \textbf{Second Stage}, which employs the built groups to select which APs are scheduled in every MAPC transmission. We consider the two stages are performed at the CC and communicated to the Sharing AP.

%, into coordinated slots. 

\subsection{First Stage: SR-Compatible group formation}

The First Stage provides a procedure to verify the compatibility of a group of APs to transmit simultaneously. We have designed an algorithm to create groups of compatible APs based on the SINR at their associated stations. 

\subsubsection{SINR requirements}

As previously indicated in Section~\ref{section:MAPC}, we assume reciprocal uplink and downlink links, so the power overheard by all the APs from uplink frames is assumed to be equivalent to the power received at the stations from downlink transmissions. All APs share this information with the CC, who stores it in a database for all AP-STA links.

%For a given subset of APs the CC verifies if they are compatible using the following expression:
For a given subset of APs, of size $M$, the power received at each of their stations must exceed the aggregate interference level by, at least, the value of $\gamma$. The CC verifies such a condition using the following expression:
\begin{align}\label{SINR_expression}\nonumber
\min_{i=1.. M}\left( 10\log_{10}(\Bar{P}^{i}) - 10\log_{10}\left(\Bar{W} + \sum_{\substack{j=1 \\ i\neq j}}^{J}{\Bar{P}^{i}_{j}} \right) \right) &\geq \gamma \\ 
\end{align}
where $\Bar{P}^{i}$ and $\Bar{P}^{i}_{j}$ are vectors that contain the RSSI values seen from all the stations associated to AP$_i$, when AP$_i$ and AP$_j$ (potential interferer) transmit at the same time.\footnote{Note that to create a group, we add APs sequentially. When a new AP is added, we guarantee both it is compatible with the rest of APs in the group, and that the other APs in the group are also compatible with the new one.} Besides, $W$ is the noise power, and $\gamma$ is the SINR threshold used. Note that, the higher the value of $\gamma$, the lower the probability of finding groups with a high number of compatible access points. Thus, all APs in the subset must satisfy (\ref{SINR_expression}) to form an SR-compatible group.

%Thus, for all values of $i$ in (\ref{SINR_expression}), the term on the left must satisfy this expression, otherwise this specific group of $J$ access points cannot form an SR-compatible group.

\subsubsection{SR-compatible group formation,  At-most-K}\label{SR-compatible-group-formation}

%For a given subset of APs, of size $J$, the power received at each of their stations must exceed the aggregate interference level by, at least, the value of $\gamma$ (see  (\ref{SINR_expression})).

In the following we describe the At-most-K algorithm we introduce in this paper to create these groups of compatible APs. 

\begin{comment}
\begin{figure*}[h]
    \centering
    \includegraphics[scale=0.85]{c-TDMA-SR_scheduling_first_stage_BF_flow.pdf}
    \caption{Flow diagram of the First Stage for c-TDMA/SR scheduling. \textcolor{red}{Remove the optional chart}}
    \label{Fig:c-TDMA-SR_scheduling_first_stage_flow}
\end{figure*}
\end{comment}

% The procedure to evaluate the spatial reuse compatibility between APs is shown in Fig.~\ref{Fig:c-TDMA-SR_scheduling_first_stage_flow} \textcolor{red}{(Update it with the proper indexes)}. It explores subset of APs, to create groups of compatible APs. Indexes $n = 1, 2, .., M$, $m = 0, 1, ..., M$ and $l =1, 2, ..., N_{cp}$ means index of the AP of reference, the index of the evaluated AP and the index of the SR-compatible group, respectively. Besides, $M$ is the total number of APs and $N_{\rm cp}$ is the total number of SR-compatible groups. 

The creation of the groups is performed taking every AP in the network as reference (or head of a group), thus resulting in a number of groups at most equal to the number of APs. Then, we add another AP at every iteration until the maximum number of APs allowed in a group, $K$, is reached.\footnote{K has been empirically selected for this paper, and a detailed analysis of this is left for future work.} For example, considering AP$_1$ as the reference (the CC is building the first coordinated group), it means that this group may contain up to the $K-1$ most convenient access points from the perspective of AP$_1$, i.e., the algorithm sequentially adds the other $K-1$ APs that have the lowest values of RSSI seen from stations associated to AP$_1$ (the highest RSSI value from the other APs seen from the stations associated to AP$_1$ covers the worst case). Thus, at every iteration, one of these pre-selected APs is added to the group only if the SINR at all the involved stations is above $\gamma$ in (\ref{SINR_expression}). Then, the operation is repeated selecting another AP as the reference, and so on. 

This condition of selecting the most compatible APs from the perspective of only one AP does not guarantee that the best possible SR group is created, because they probably are not the best choice for other APs in the group different from the reference one.
%, but it often means the highest SINR value for stations associated to the reference AP, AP$_1$ in the example. 
Therefore, selecting a value of $\gamma$ high enough is crucial to guarantee good SINR levels for all stations associated to any of the APs that belong to the same group, even if they are not the reference one.

Note that, better-located APs will belong to several groups, but the final decision about which group(s) will transmit in the next TXOP is done in the Second Stage and it depends on the number of packets that APs have in their buffers, which is directly related to the time that a group has been waiting for transmitting. 
% The main benefit of the proposed mechanism is that an AP can belong to several groups, which can also be advantageous for the Second Stage (each AP in several groups means more flexibility).

\subsection{Second Stage: Traffic Scheduling Algorithms}

The Second Stage is intended for scheduling one SR-compatible group (from the ones previously computed in the First Stage) per coordinated slot based on the buffer state information collected from all APs, which contains the number of packets in the transmission buffer of each AP, as well as the arrival time of the oldest packet. 

\paragraph{NumPkSingle} The CC selects the AP with the highest number of packets waiting in the buffer. Then, considering only the groups created in the First Stage that contain the selected AP, the CC schedules the group of APs with the highest number of packets, calculated as the sum across all individual APs.

%contribution (the sum of the number of packets for a single station of all APs in the group).

\paragraph{NumPkGroup} The CC selects the group of APs with the highest number of packets.

\paragraph{OldPkSingle} The CC selects the AP with the oldest packet waiting in the buffer. Then using the groups created in the First Stage that contain the selected AP, the CC schedules the group with the highest aggregate delay, calculated as the sum of the waiting time of the oldest packet across all individual APs.

%contribution (the sum of the time that oldest packreet in the buffer of each individual AP has been waiting).

\paragraph{OldPkGroup} The CC selects the group with the maximum aggregate group delay, i.e., the sum of the time that the oldest packet in the buffer of each individual AP has been waiting.

For the sake of fairness between groups, NumPkGroup (OldPkGroup) values are normalized by the number of APs in each group, so we avoid that groups with a few (large) number of APs starve.

% ----------------------------------_
% ----------------------------------_
% ----------------------------------_
% ----------------------------------_

\section{System Model}\label{system_model}

To evaluate the performance of the MAPC framework, and assess the operation of the group creation algorithm and the proposed traffic schedulers, we consider the following scenario.

\subsection{Deployment}

%We consider a WLAN that consists of $M=9$ Basic Service Sets (BSSs) where one access point and several stations have been  deployed in each of them, following a grid topology as shown in Fig.~\ref{Fig:single_scenario}. 

%We consider a WLAN that consists of $M=9$ Basic Service Sets (BSSs) where one access point and several stations have been  deployed in each of them, following a grid topology as shown in Fig.~\ref{Fig:single_scenario}. 

\begin{figure}[th]
    \centering
    \includegraphics[scale=0.62]{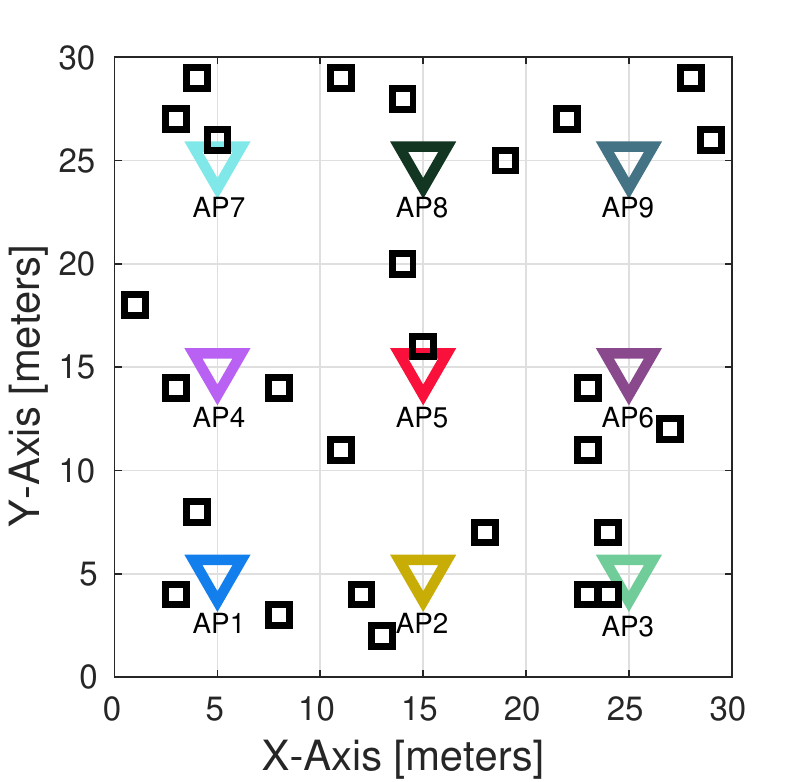}
    \caption{An Enterprise WLAN scenario, with multiple APs.}
    \label{Fig:single_scenario}
\end{figure}

We divide the area of interest in $9$ subareas of 10x10 meters each, and at the center of each subarea we deploy one AP as shown in Fig.~\ref{Fig:single_scenario}. All APs are set to operate in the same channel. We designate AP$_5$ (the AP in the middle) as the Sharing AP since all the other APs are within its coverage area. $N=3$ stations are randomly placed in each subarea, and associated to the nearest AP, which in this case always corresponds to the one placed at the subarea center. 

%Given the size of the area of interest, not all APs are within the coverage area of the others. Therefore, we fix AP$_5$ (the AP in the middle) as the Sharing AP so all other APs can correctly receive and decode any transmitted frame in the WLAN. $N=3$ stations are randomly placed in each subarea, and associated to the nearest AP, which in this case always corresponds to the one placed at its center. 

Multiple transmission rates are allowed, so stations close (far) from their AP use higher (lower) MCSs to transmit and receive data. The MCS used by a given AP to transmit to a station depends on the SINR observed by the station. This value is estimated by the CC given the group of APs that will simultaneously transmit using the RSSI information collected from uplink frames. This value is announced in the MAP-TF frames by the sharing AP when a coordinated transmission starts. To allocate a specific MCS to an AP-STA pair, we employ the curves presented in \cite{EvgenyMCS}. These curves give the SINR ranges corresponding to each MCS that ensure an error-free transmission. A-MPDU transmissions are enabled in each coordinated slot, with the maximum number of aggregated packets depending on the MCS used and the slot duration.

%The area of interest is divided in $9$ subareas of 10x10 meters each, deploying the corresponding AP at the center of each one. We set all APs to operate in the same channel, and we assume they are within the coverage area of the others. Therefore all of them can correctly receive and decode any transmitted frame in the WLAN. $N=3$ stations are randomly placed in each subarea, and associated to the nearest AP. Multiple transmission rates are allowed, so stations close (far) from their AP use higher (lower) MCSs to transmit and receive data. The MCS used by a given AP to transmit to a certain station depends on the SINR at the receiver (which is estimated by CC for each group of simultaneous transmitting APs using the RSSI information collected from uplink frames), \hl{and announced by the sharing AP when a coordinated transmission starts}. To allocate a specific MCS to a station, we employ the curves presented in \cite{EvgenyMCS}, which define the SINR ranges corresponding to each MCS to ensure an error-free transmission.

%Both the clear channel assessment (CCA) / preamble detection (PD) threshold and the energy detection (ED) threshold are set to $-82$ dBms. 

%Each of them has been placed in the center of a subarea of $A$ m$^2$. 

%Our system consists of several access points deployed in a building. For the sake of simplicity, we consider only $M$ of them, allocated in the same channel (which means they are overlapping) and distributed over the entire building, see Fig.~\ref{Fig:single_scenario}. We assume they are within the coverage area of the others, and therefore all of them can correctly receive and decode any transmitted frame in the WLAN. Each of them has been placed in the center of a subarea of $A$ m$^2$. 

% Thus, in the case of c-TDMA/SR transmissions, the CC should successfully find a subset of those APs that do not cause unacceptable interference to each other. We assume that the Sharing AP is placed at the center of the area, so it can be overheard by the rest of the devices.

The path loss is modelled using the TGax model for Enterprise Scenarios \cite{pathloss}:
\begin{equation}\label{pl_equation}
P_{L} = 40.05 + 20\log_{10}\left(\frac{\min(d,B_{p})f_{c}}{2.4}\right) + P' + 7W_{n} \text{,}
\end{equation}
where $d$ is the distance between the transmitter and the receiver in meters, $f_{c}$ is the central frequency in GHz, $W_{n}$ is the number of walls and $P'$ is given by $P' = 35\log_{10}(d/B_{p})$,
when $d$ is higher than the breaking point $B_{p}$. Otherwise, it is zero.

% \begin{figure*}[t!!]
%     \centering
%     \includegraphics[scale=0.65]{tx_model.pdf}
%     \caption{Proposed model for best effort traffic. \bcom{We need to find better and shorter names for the TXOP-sharing and TXOP-MAX. Indeed, the TXOP-MAX should be TXOP-sharing-MAX}}.
%     \label{Fig:model}
% \end{figure*}

Only downlink traffic is considered, i.e., from the AP to the stations. The traffic generation process works as follows: Just before every MAPC transmission, $N_{p}=10$ packets for each station arrive at its corresponding AP with probability $p$ depending on the considered traffic load. Three different per-STA load levels are employed to refer to low, medium and high load conditions: $1$, $6$ and $8$ Mbps, respectively. For the sake of simplicity, we assume that buffers in all APs are large enough so that incoming packets are never lost due to buffer overflow.

\subsection{Operation}

The proposed MAPC framework operates as it is shown in Fig.~\ref{Fig:c-TDMA/SR}. MAPC transmissions are scheduled to start at every $nT, n = 0, 1, \ldots, N_{\rm SN}-1$, where $N_{\rm SN}=10000$ is the total number of MAPC transmissions considered in each simulation. Each MAPC transmission consists of several c-TDMA or c-TDMA/SR length-variable coordinated slots (the actual value depends on the number of packets and the MCSs used to transmit them), followed by uncoordinated transmissions using the legacy CSMA/CA mechanism. The duration of a given MAPC transmission is determined by the scheduler, although it must not exceed its maximum value, i.e., $T_{\rm TXOP-MAX}$. 
%Considering the requirements of the traffic flowing towards the stations, the CC can modify the value of $T$ to cover the demand in terms of delay for the most sensitive cases.
The $T_{\rm TXOP-MAX}$ duration is also a parameter used by the CC to control the amount of time devoted to MAPC transmissions, and it can be modified depending on the scenario.

\begin{figure*}[t]
    \begin{subfigure}[b]{0.325\textwidth}
        \centering
        \includegraphics[width=\textwidth]{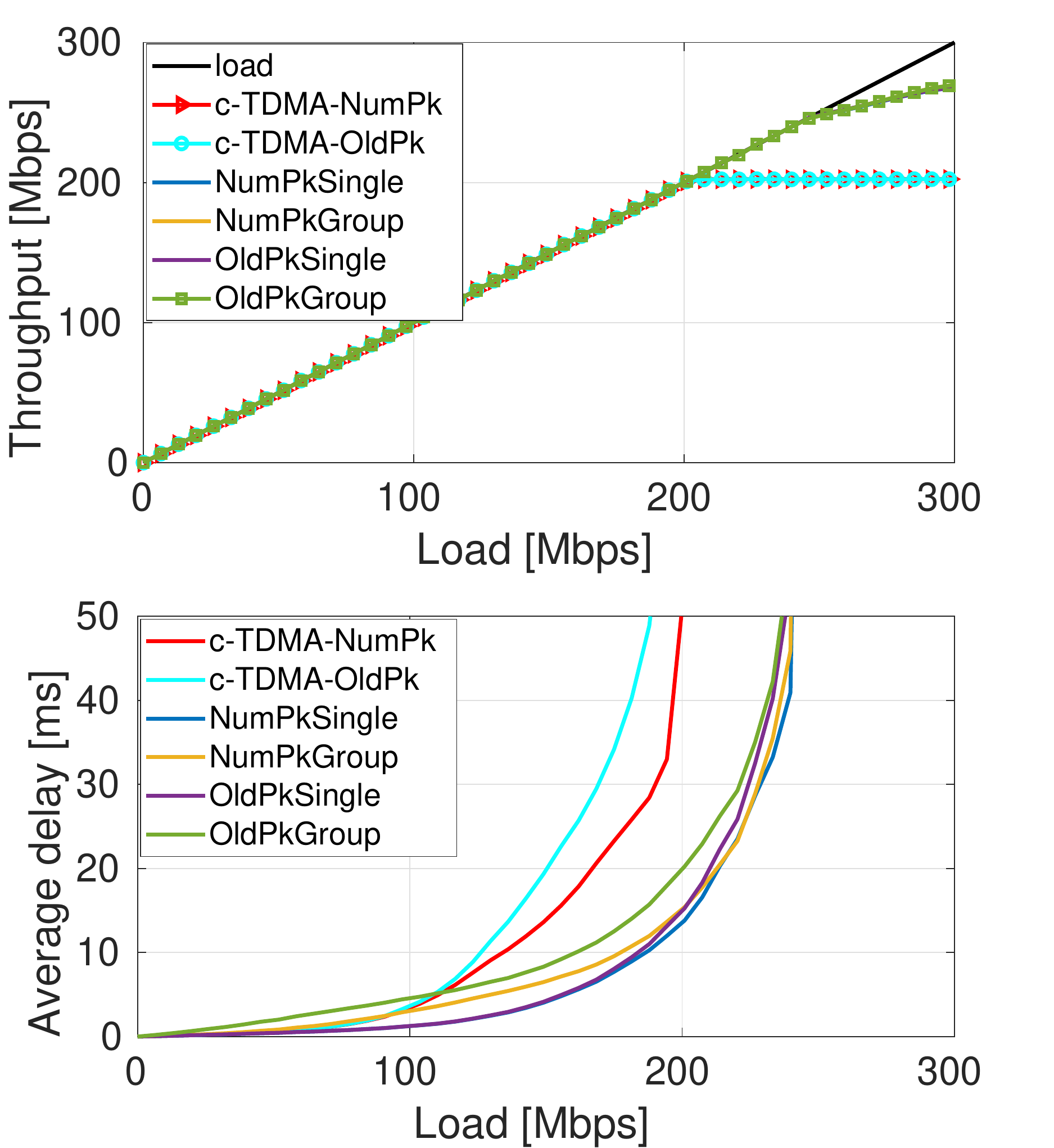}
        \caption{$\gamma = 20$ dB and $K = 3$.}
        \label{Fig:toy1_results}
    \end{subfigure}
    \hfill
    \begin{subfigure}[b]{0.325\textwidth}
        \centering
        \includegraphics[width=\textwidth]{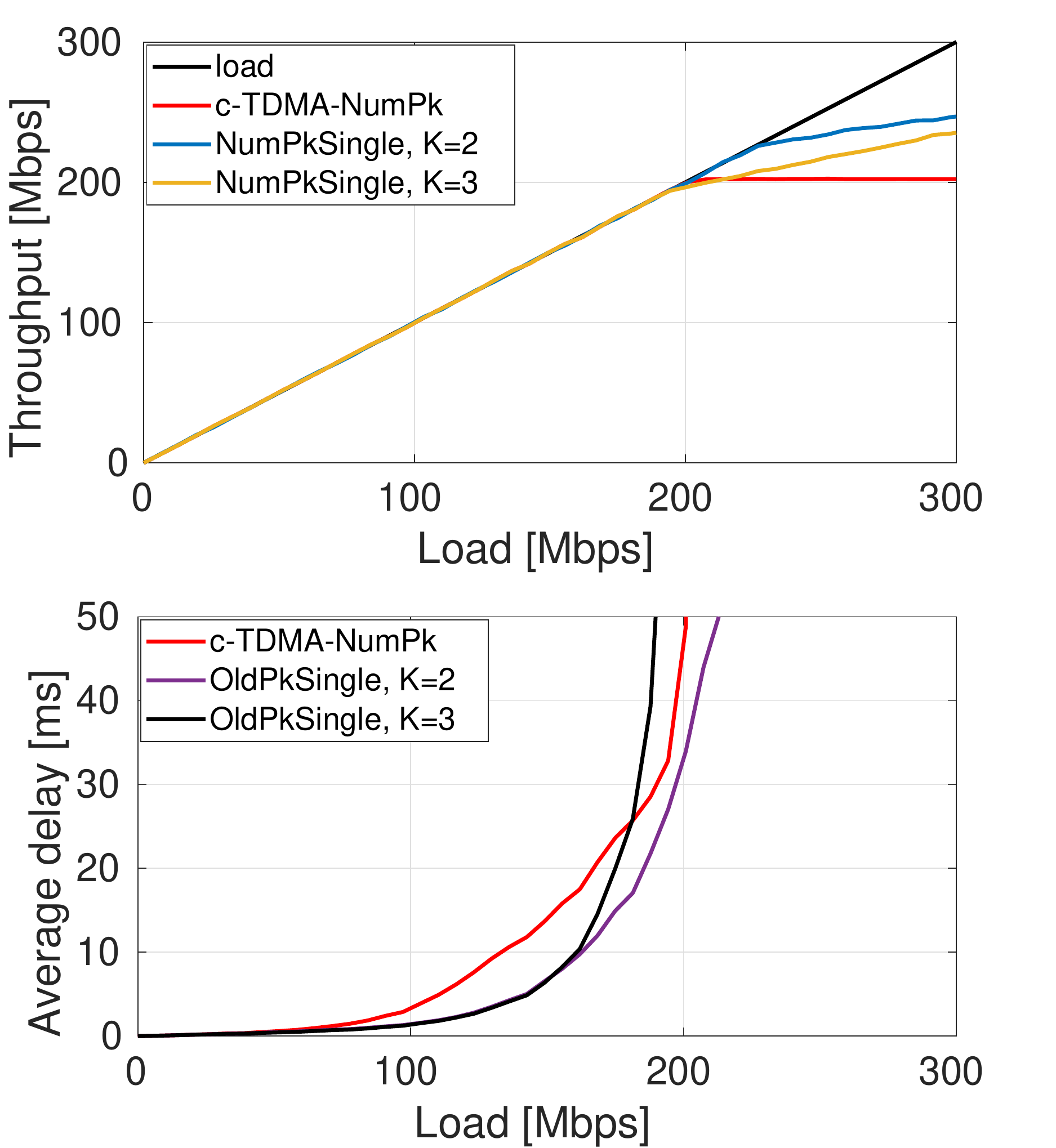}
        \caption{$K=2$ and $3$, with $\gamma = 14$ dB.}
        \label{Fig:toy2_results_gamma_14}
    \end{subfigure}
    \hfill
    \begin{subfigure}[b]{0.325\textwidth}
        \centering
        \includegraphics[width=\textwidth]{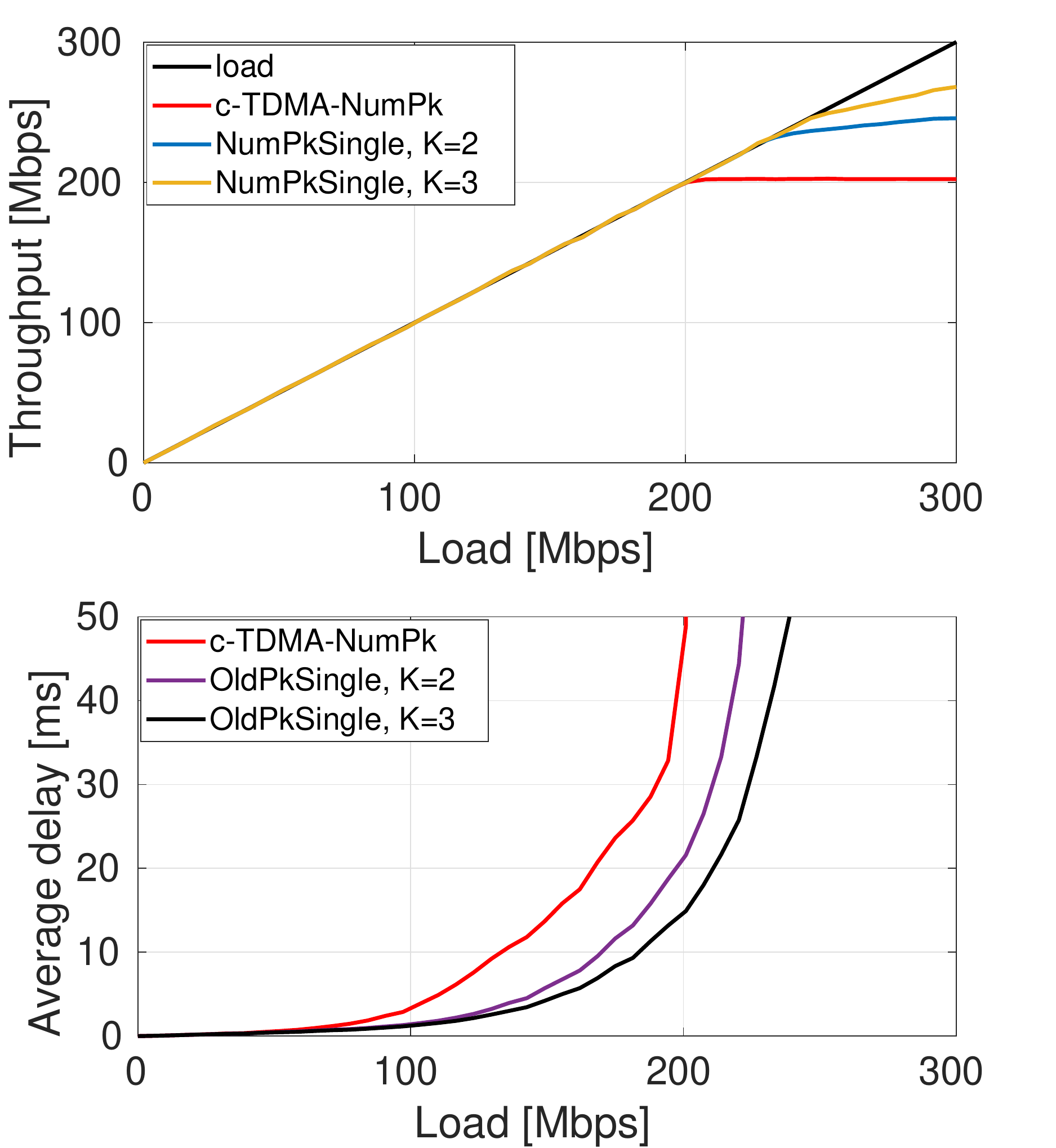}
        \caption{$K=2$ and $3$, with $\gamma = 20$ dB.}
        \label{Fig:toy2_results_gamma_20}
    \end{subfigure}
    \caption{Aggregate (WLAN) throughput and average delay in the scenario of Fig. \ref{Fig:single_scenario} for different values of $\gamma$ and $K$.}
\end{figure*}

\section{Simulation Results}

We present and discuss the results obtained from running simulations on Matlab and taking the Scenario described in Section \ref{system_model}. Results labelled with c-TDMA (without SR) correspond to the case where only one AP is allowed to transmit in each coordinated slot of a MAPC transmission.\footnote{The advantages of c-TDMA/SR vs legacy 802.11 are shown in \cite{TXOPsharingPaper}.} The parameters used for the numerical simulations are shown in Table~\ref{tab2_simulation_paramters}.

\begin{table}
    \caption{Simulation Parameters.}
    \label{tab2_simulation_paramters}
    \begin{center}
        \begin{tabular}{|c|c|c|c|}
        \hline
        \textbf{Parameter} & \textbf{Value} & \textbf{Parameter} & \textbf{Value} \\
        \hline
        $B_{p}$ &  10 & $W_{n}$ &  3 \\
        \hline
        Number of spatial streams & 1  & $T$ [ms] & 5 \\
        \hline
        $T_{\rm TXOP-MAX}$ [ms] & 3 & 11ax MCS [index] & 0-10  \\
        \hline
        $T_{\text{MAP-RTS}}$ [$\mu$s] & 80 & $T_{\text{MAP-CTS}}$ [$\mu$s] & 62 \\
        \hline
        $T_{\rm CTS-TO}$ [$\mu$s] & 41 & $T_{\rm MAP-TF}$ [$\mu$s] &  76\\
        \hline
        T$_{e}$ [$\mu$s] &  9 & AP Tx-power [dBm] & 23 \\
        \hline
        $L$ [bytes] & 1500 & CCA [dBm] & -82 \\
        \hline
        \end{tabular}
    \end{center}
\end{table}

\subsection{A specific scenario}

% \begin{figure}[t!!!!]
    %\hspace{20mm}
    % \begin{subfigure}[b]{0.24\textwidth}
    %     \centering
    %     % \includegraphics[scale=0.5]{toy-1.eps}
    %     \includegraphics[width=\textwidth]{toy-1.eps}
    %     \caption{Toy scenario 1.}
    %     \label{Fig:toy1}
    % \end{subfigure}
    %\hspace{30mm}
% \begin{figure}[t!]
%         \centering
%         \includegraphics[scale=0.7]{toy-2.eps}
%         % \includegraphics[width=\textwidth]{toy-2.eps}
%         \caption{Toy scenario 2.}
%         \label{Fig:toy2}
%     \caption{Single scenario. 9 APs and 27 stations have been placed in the entire building. In each 10x10 m$^2$ room, one AP located at the center and 3 stations uniformly deployed at random.}
% \end{figure}
% \end{figure}

\begin{figure}[h!!]
    \centering
    \includegraphics[scale=0.38]{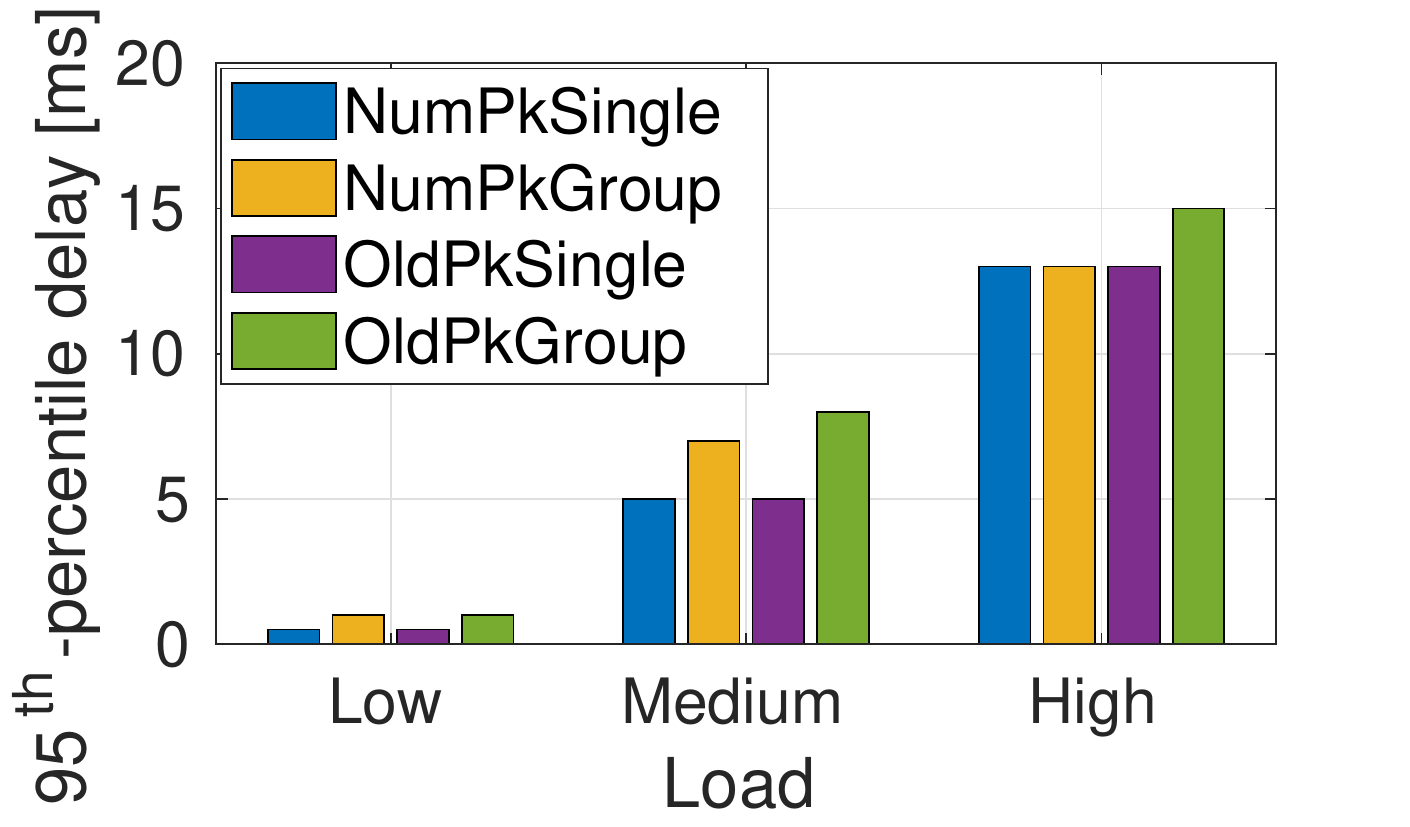}
    \caption{95$^{th}$-percentile delay for different algorithms in the scenario of Fig. \ref{Fig:single_scenario}, $\gamma = 20$ and $K=3$.}
    \label{Fig:toy2_95perctile_delay}
\end{figure}

In this section, we study the scenario shown in Fig.~\ref{Fig:single_scenario} with nine APs and three STAs per AP. Fig.~\ref{Fig:toy1_results} shows the aggregate throughput and average delay, using NumPkSingle, NumPkGroup, OldPkSingle, OldPkGroup, c-TDMA-NumPk and c-TDMA-OldPk algorithms. The same value of $\gamma = 20~$dBs is used in all cases, as well as the maximum number of APs per group, i.e., $K=3$, for the SR-compatible group formation. With respect to the throughput, as expected, c-TDMA approaches reach saturation before the SR ones. 
%\bcom{Not clear what is low, medium and high load conditions. What is the load used for Fig.\ref{Fig:single_scenario}?}\dcom{The load in Fig. 3? It is a sweep from 0-max load} \dcom{We also defined specific loads, for the cases where only one value of load was used, resulting in low=1, medium=6, high=8 Mbps. } 
In the case of the average delay, similarly, SR approaches outperform c-TDMA ones. Moreover, comparing SR scheduling algorithms between them, NumPkSingle and OldPkSingle perform better than the rest on low ($<100$ Mbps) and medium (100-200 Mbps) load conditions, pointing out the advantage of always scheduling the AP with more packets in the buffer. For high load all the algorithms that support SR transmissions perform similar.
%, showing a difference over $15$ ms \bcom{where do you see this?}\dcom{Only against the c-TDMA ones, and 2-5 ms against NumPkGroup and OldPKGroup} with respect to c-TDMA-OldPk on medium load conditions. 

%The difference between the scheduling algorithms based on single AP and the ones based on group selection (easily observed for low and medium load) is due to the condition imposed to normalize the groups to avoid starvation in the group-based algorithms. This normalization sometimes results in an inefficient scheduling decision because the CC does not select the group with a highest number of packets (delay). Therefore, there is room to improve per-Group scheduling algorithms in future works.  

%most packets/delay waiting in the buffers. Therefore, there is room to improve per-Group scheduling algorithms in future works.  

Figs. \ref{Fig:toy2_results_gamma_14} and \ref{Fig:toy2_results_gamma_20} show the aggregate throughput and average delay for different values of $\gamma$ and $K$. The case when $\gamma=14$ is used (Fig. \ref{Fig:toy2_results_gamma_14}), exhibits a worse performance (saturation throughput is reached earlier) for $K=3$ than for $K=2$, due to using a low value of $\gamma$. The reason is that in some cases the SINR at the stations with $K=3$ is lower than for $K=2$, and therefore a more robust MCS index will be used for $K=3$. On the contrary, Fig.~\ref{Fig:toy2_results_gamma_20} exhibits a gain for $K=3$ with respect to $K=2$ because the value of $\gamma$ is set to guarantee that the SINR at the stations is at least $20$ dB. Thus, even if we have less groups with three APs, the high value of $\gamma$ guarantees they will be able to use high MCS for the data transmission. In summary, both $\gamma$ and $K$ are tuneable parameters that can be optimized for each individual scenario.
%some of the groups now may consist of three APs using also higher MCS indexes (due to have $\gamma=20$ dB, which is four dBs higher than the previous case), so it increases the overall performance of their transmissions.

Fig.~\ref{Fig:toy2_95perctile_delay} shows the 95$^{th}$-percentile delay achieved by the different algorithms for the case with $K=3$ and $\gamma=20$. In all cases, per-AP schemes outperform per-Group ones. Interestingly, scheduling algorithms based on the number of packets in the buffer outperform delay based ones also in the worst-case delay since they are able to schedule more efficient transmissions.

%In low load conditions (1 Mbps per station), all the algorithms perform similar (delay lower than 1 ms), except NumPkGroup and OldPkGroup which show a slightly higher delay. For medium (6 Mbps per station) load, NumPkSingle and OldPkSingle still perform better than the rest, having a noticeable difference with OldPkGroup (around $3$ ms) and NumPKGroup ($2$ ms) due to the aforementioned negative affect of normalizing the groups. Finally, OldPkGroup is again the worst for high load, having a difference of $2$ ms when compared with the rest of algorithms.

\subsection{Random scenarios}

Results in this section are obtained through the simulation of 1000 random deployments. In each deployment, while stations are generated uniformly at random inside each subarea, APs are kept at the center of each one. In all scenarios, we have considered $K=3$, $\gamma=20$, and a traffic load of $8$ Mbps per station, which represents an aggregate load of $216$ Mbps.

\begin{figure}
        \centering
        \includegraphics[scale=0.38]{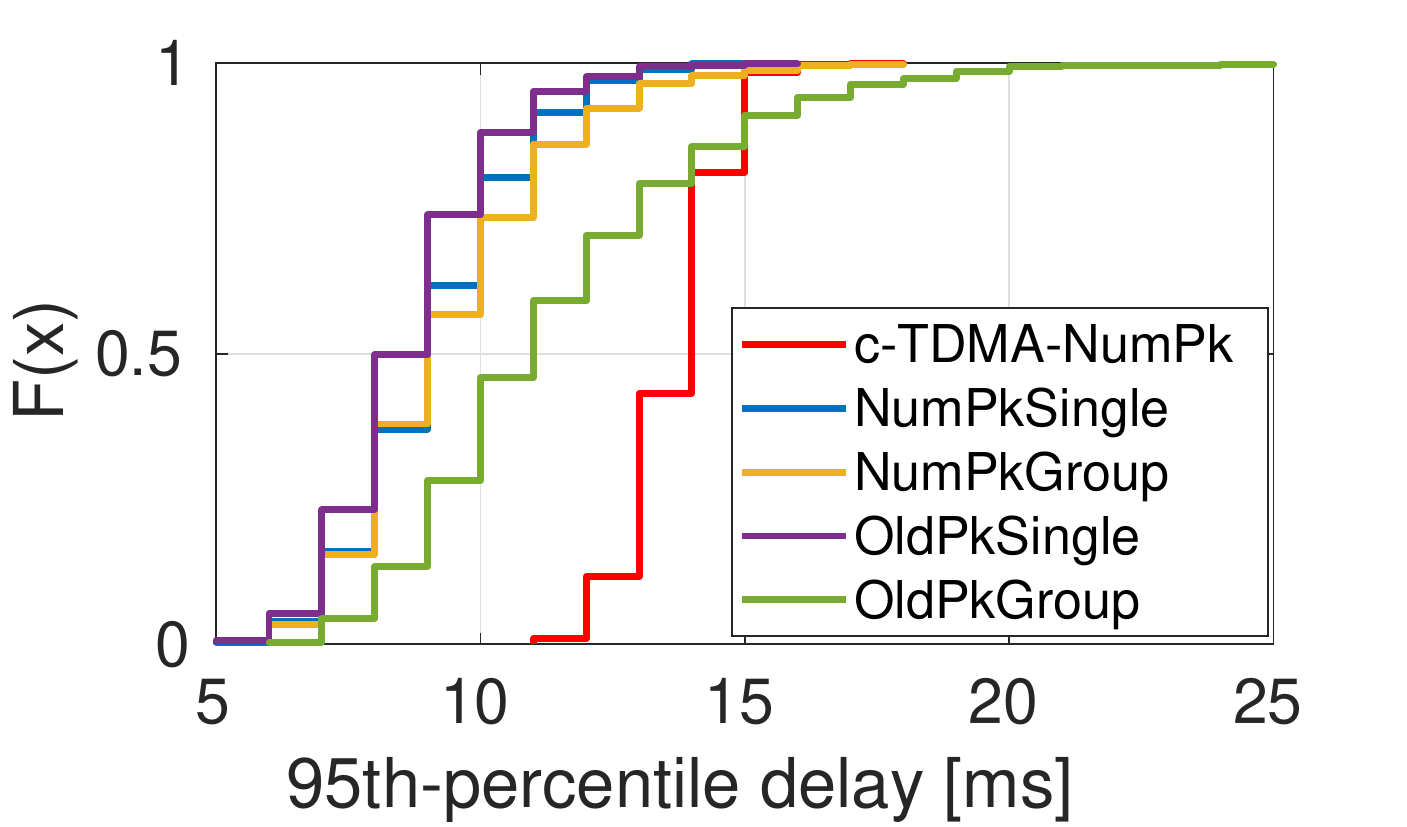}
        \caption{CDF of the 95$^{th}$-percentile delay between random scenarios. %The SR-compatibility between APs has been obtained with $\gamma = 20$ and at most 2 access points per group, i.e., $K=3$.
        }
        \label{Fig:random_95perctile_delay_high_load}
\end{figure}

Fig.~\ref{Fig:random_95perctile_delay_high_load} exhibits the Cumulative Distribution Function (CDF) of the 95$^{th}$-percentile delay over multiple generated random scenarios. c-TDMA algorithms show the worst delay, with a difference between them and c-SR ones exceeding $4$ ms for most of the percentiles. Note that, as before, the algorithms that schedule the groups based on the number of packets, perform better than the ones using the oldest packet criterion.  The reason is that the delay-based algorithms are not able to schedule as many packets per MAPC transmission as the number of packets-based ones, which turns out to be also counterproductive in terms of delay. 

%fail to achieve that because they are not in general able to schedule as many packets are the ones based on the buffer state, which turns out to have a negative impact on the overall network performance. 

%since they in average schedule less packets per MAPC transmission, thus leaving unused, and resulting in a lower overall performance.

%are not able to match the ones based on the number of packets with respect to their ability to use efficiently the MAPC transmission by filling 

%less efficient than the ones based on the number of packets with respect to their ability to use 

%number of packets that are transmitter per MAPC-transmission, resulting in a waste of time reserved for MAPC.

%are not efficient enough to fill with packets the entire $T_{\rm TXOP-MAX}$ as the ones based on the number of packets, resulting in a waste of time reserved for MAPC.

\begin{figure}
        \centering
        \includegraphics[scale=0.38]{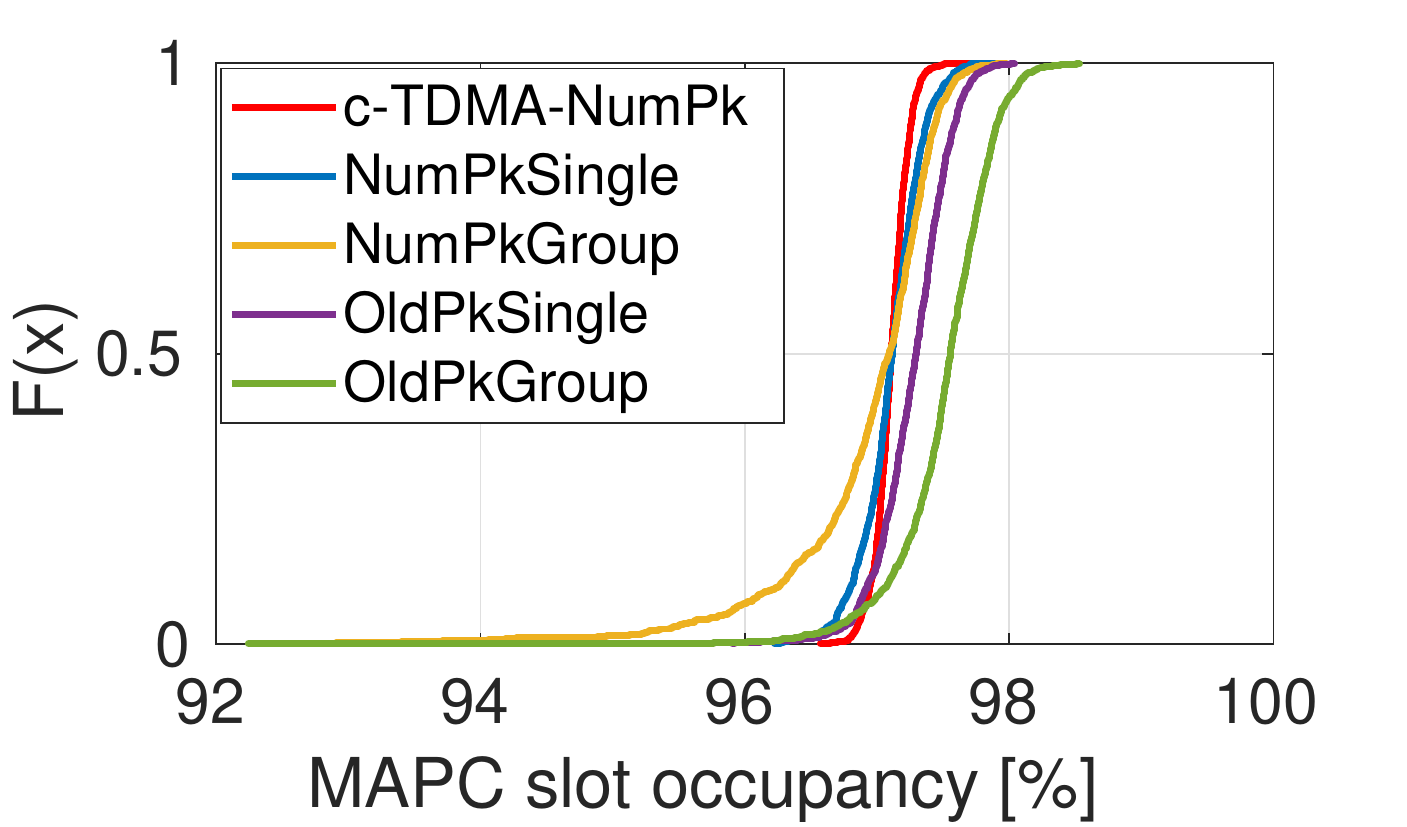}
    \caption{CDF of the MAPC slot occupancy  between random scenarios. %Results have been obtained for $\gamma = 20$ and $K=3$.
    }
    \label{Fig:slot_occupancy}
\end{figure}

Finally, Fig.~\ref{Fig:slot_occupancy} shows the MAPC slot occupancy, defined as the ratio between the time spent by cooperative transmissions and the maximum duration of MAPC transmission. The lower the ratio, the more efficient the use of the MAPC slots, increasing the amount of time available for uncoordinated CSMA/CA transmissions. Note that, OldPkGroup not only achieves the worst overall results as seen previously, but also it requires the largest MAPC transmissions.

%     \caption{CDF of the 95$^{th}$-percentile delay of random scenarios for different algorithms in medium and high load conditions.}
% \end{figure*}

% \begin{figure}[t!]
%    \begin{subfigure}[b]{0.3\textwidth}
%        \centering
%        \includegraphics[width=\textwidth]{MAPC_slot_occupancy_low_load.eps}
%        \caption{Low load.}
%        \label{Fig:MAPC_slot_occupancy_low_load}
%    \end{subfigure}
%    \begin{subfigure}[b]{0.3\textwidth}
%        \centering
%        \includegraphics[width=\textwidth]{MAPC_slot_occupancy_medium_load.eps}
%        \caption{Medium load.}
%        \label{Fig:MAPC_slot_occupancy_medium_load}
%    \end{subfigure}

\section{Conclusions}

In this paper, we have introduced and evaluated a framework for multi-AP coordinated transmissions. We also proposed a method to create groups of SR-compatible APs, and a set of algorithms to schedule coordinated transmissions. Results evidence that algorithms based-on per-AP selection perform better than the per-Group selection ones.

\bibliographystyle{IEEEtran}
\bibliography{main}

\begin{thebibliography}{10}
\providecommand{\url}[1]{#1}
\csname url@samestyle\endcsname
\providecommand{\newblock}{\relax}
\providecommand{\bibinfo}[2]{#2}
\providecommand{\BIBentrySTDinterwordspacing}{\spaceskip=0pt\relax}
\providecommand{\BIBentryALTinterwordstretchfactor}{4}
\providecommand{\BIBentryALTinterwordspacing}{\spaceskip=\fontdimen2\font plus
\BIBentryALTinterwordstretchfactor\fontdimen3\font minus
  \fontdimen4\font\relax}
\providecommand{\BIBforeignlanguage}[2]{{%
\expandafter\ifx\csname l@#1\endcsname\relax
\typeout{** WARNING: IEEEtran.bst: No hyphenation pattern has been}%
\typeout{** loaded for the language `#1'. Using the pattern for}%
\typeout{** the default language instead.}%
\else
\language=\csname l@#1\endcsname
\fi
#2}}
\providecommand{\BIBdecl}{\relax}
\BIBdecl

\bibitem{Internet_traffic_Sandvine}
Sandvine, ``{The Global Internet Phenomena Report},''
  \url{https://www.sandvine.com/phenomena}, 2022.

\bibitem{carrascosa2022cloud}
M.~Carrascosa and B.~Bellalta, ``{Cloud-gaming: Analysis of google stadia
  traffic},'' \emph{Computer Communications}, vol. 188, pp. 99--116, 2022.

\bibitem{zhao2021virtual}
S.~Zhao, H.~Abou-zeid, R.~Atawia, Y.~S.~K. Manjunath, A.~B. Sediq, and X.-P.
  Zhang, ``{Virtual Reality Gaming on the Cloud: A Reality Check},''
  \emph{arXiv preprint arXiv:2109.10114}, 2021.

\bibitem{UHRobjectives}
``{IEEE 802.11-22/0078r3: 802.11 UHR Draft Proposed PAR},''
  \url{https://mentor.ieee.org/802.11/dcn/23/11-23-0078-03-0uhr-uhr-draft-proposed-par.docx},
  January 2023, accessed on 08/02/2023.

\bibitem{AdrianGarciaSurvey}
A.~Garcia-Rodriguez, D.~López-Pérez, L.~Galati-Giordano, and G.~Geraci,
  ``{IEEE 802.11be: Wi-Fi 7 Strikes Back},'' \emph{IEEE Communications
  Magazine}, vol.~59, no.~4, pp. 102--108, 2021.

\bibitem{EvgenySurvey}
E.~Khorov, I.~Levitsky, and I.~F. Akyildiz, ``{Current Status and Directions of
  IEEE 802.11be, the Future Wi-Fi 7},'' \emph{IEEE Access}, vol.~8, pp.
  88\,664--88\,688, 2020.

\bibitem{TXOPsharingPaper}
D.~Nunez, F.~Wilhelmi, S.~Avallone, M.~Smith, and B.~Bellalta, ``{TXOP sharing
  with Coordinated Spatial Reuse in Multi-AP Cooperative IEEE 802.11be
  WLANs},'' in \emph{2022 IEEE 19th Annual Consumer Communications Networking
  Conference (CCNC)}, 2022, pp. 864--870.

\bibitem{draft11be}
``{IEEE P802.11be Draft Standard for Information technology—
  Telecommunications and information exchange between systems Local and
  metropolitan area networks— Specific requirements. Part 11: Wireless LAN
  Medium Access Control (MAC) and Physical Layer (PHY) Specifications.
  Amendment 8: Enhancements for extremely high throughput (EHT)},'' May 2022.

\bibitem{lopez2022multi}
{\'A}.~L{\'o}pez-Ravent{\'o}s and B.~Bellalta, ``{Multi-link Operation in IEEE
  802.11 be WLANs},'' \emph{IEEE Wireless Communications (arXiv preprint
  arXiv:2201.07499)}, 2022.

\bibitem{carrascosa2022experimental}
M.~Carrascosa, G.~Geraci, E.~Knightly, and B.~Bellalta, ``{An experimental
  study of latency for IEEE 802.11 be multi-link operation},'' in \emph{ICC
  2022-IEEE International Conference on Communications}.\hskip 1em plus 0.5em
  minus 0.4em\relax IEEE, 2022, pp. 2507--2512.

\bibitem{MentorConsiderations0590r5}
J.~Han \emph{et~al.}, ``{Coordinated Spatial Reuse: Focus on Downlink},'' May
  2020, doc.: IEEE 802.11-20/0590r5.

\bibitem{MentorSR1534r1}
K.~Aio \emph{et~al.}, ``{Coordinated Spatial Reuse Performance Analysis},''
  September 2019, doc.: IEEE 802.11-19/1534r5.

\bibitem{MentorSR0107r1}
D.~Akhmetov \emph{et~al.}, ``{Multi-AP coordination for spatial reuse},'' 2020,
  doc.: IEEE 802.11-20/0107r1.

\bibitem{MentorSR0576r1}
Y.~Seok \emph{et~al.}, ``{Coordinated Spatial Reuse (C-SR) Protocol},'' April
  2020, doc.: IEEE 802.11-20/0576r1.

\bibitem{WoojinCoOFDMAframework}
\BIBentryALTinterwordspacing
W.~Ahn, ``{Novel Multi-AP Coordinated Transmission Scheme for 7th Generation
  WLAN 802.11be},'' \emph{Entropy}, vol.~22, no.~12, 2020. [Online]. Available:
  \url{https://www.mdpi.com/1099-4300/22/12/1426}
\BIBentrySTDinterwordspacing

\bibitem{Yuto_kihira_cSR_Qlearning}
Y.~Kihira, K.~Yamamoto, A.~Taya, T.~Nishio, Y.~Koda, and K.~Yano,
  ``Interference-free ap identification and shared information reduction for
  tabular q-learning-based wlan coordinated spatial reuse,'' \emph{IEICE
  Communications Express}, vol. advpub, p. 2022XBL0051, 2022.

\bibitem{EvgenyMCS}
A.~Krotov, A.~Kiryanov, and E.~Khorov, ``{Rate Control With Spatial Reuse for
  Wi-Fi 6 Dense Deployments},'' \emph{IEEE Access}, vol.~8, pp.
  168\,898--168\,909, 2020.

\bibitem{pathloss}
S.~Merlin \emph{et~al.}, ``{TGax Simulation Scenarios},'' Nov. 2015, doc.: IEEE
  802.11-14/0980r16.

\end{thebibliography}

\end{document}